\documentclass[10pt,twocolumn,letterpaper]{article}

\usepackage[pagenumbers]{cvpr} 

\usepackage{graphicx}
\usepackage{amsmath}
\usepackage{amssymb}
\usepackage{booktabs}

\usepackage[pagebackref,breaklinks,colorlinks]{hyperref}
\newcommand{\mysection}[1]{\vspace{2pt}\noindent\textbf{#1}}

\usepackage[capitalize]{cleveref}
\crefname{section}{Sec.}{Secs.}
\Crefname{section}{Section}{Sections}
\Crefname{table}{Table}{Tables}
\crefname{table}{Tab.}{Tabs.}

\def\cvprPaperID{23} 
\def\confName{CVPR}
\def\confYear{2023}

\begin{document}

\title{AgroTIC: Bridging the gap between farmers, agronomists, and merchants through smartphones and machine learning}

\author{Carlos Hinojosa\\
{\tt\small carlos.hinojosa@saber.uis.edu.co}
\and
Karen Sanchez\\
{\tt\small karen.sanchez2@saber.uis.edu.co}
\and
\hspace{-0.4in}Ariolfo Camacho\\
{\tt\small \hspace{-0.4in} ariolfo.camacho@saber.uis.edu.co}
\and
\hspace{0.3in}Henry Arguello\\
{\tt\small \hspace{0.3in} henarfu@uis.edu.co \hfill }
}
\maketitle

\begin{abstract}
   In recent years, fast technological advancements have led to the development of high-quality software and hardware, revolutionizing various industries such as the economy, health, industry, and agriculture. Specifically, applying information and communication technology (ICT) tools and the Internet of Things (IoT) in agriculture has improved productivity through sustainable food cultivation and environment preservation via efficient use of land and knowledge. However, limited access, high costs, and lack of training have created a considerable gap between farmers and ICT tools in some countries, e.g., Colombia. To address these challenges, we present AgroTIC, a smartphone-based application for agriculture that bridges the gap between farmers, agronomists, and merchants via ubiquitous technology and low-cost smartphones. AgroTIC enables farmers to monitor their crop health with the assistance of agronomists, image processing, and deep learning. Furthermore, when farmers are ready to market their agricultural products, AgroTIC provides a platform to connect them with merchants. We present a case study of the AgroTIC app among citrus fruit farmers from the Santander department in Colombia. Our study included over 200 farmers from more than 130 farms, and AgroTIC positively impacted their crop quality and production. The AgroTIC app was downloaded over 120 times during the study, and more than 170 farmers, agronomists, and merchants actively used the application.
\end{abstract}


\section{Introduction}
\label{sec:introduction}
Ensuring food security has been a global concern throughout human history. The global food crisis of 2007-2008 emphasized the importance of increasing both the quantity and quality of food production \cite{sasson2012food}. As agriculture is the primary food supply source, new and modern methods are required to ensure the world's food security. One key idea for addressing such challenges in agriculture is the use of technology, information, and communication (ICT)\cite{chhachhar2014impact, ali2016impact} tools to measure or monitor field and crop conditions such that farmers can make informed decisions at each stage of their farming process \cite{anurag2008agro,gebbers2010precision}.

However, there is a gap between farmers and ICT tools in some countries that remains remarkable due to high costs, limited access, and poor/lack of training. Colombia is a South American country known for its mountainous terrain, rainfall patterns, topography, and climate diversity, allowing the cultivation of various crops. Nevertheless, Colombian agriculture principally employs traditional methods, and most of the population with agricultural vocation has little knowledge about the advantages of using ICT. For instance, in Simacota, a town in the Santander department, from 132 sampled farming families, only 87 have smartphones, of which 11 are top-end, 20 are mid-range, and 56 are low-end. Besides, 50 families have ever used Apps like WhatsApp, 26 have ever surfed the internet, and only 15 have ever used E-mail. This example reveals the digital breach in most of the farming population in Colombia. In fact, the use of ICT in the Colombian agricultural sector faces many problems like expensive ICT equipment \cite{carulla}, the unknown advantages for the farmers, user rejection to use new technology, and poor internet coverage in rural areas. This limits the technical development and effective and efficient exploitation of food resources in different regions of Colombia.

The use of ICT and internet of things (IoT) tools can potentially increase agriculture's efficiency, productivity, and sustainability by providing timely information, knowledge sharing, and supporting decision-making \cite{muangprathub2019iot, wishkerman2017application, chung2018smartphone, TRIPATHI2020183}. For instance, authors in \cite{muangprathub2019iot} propose to develop a smartphone and web application for optimally watering agricultural crops based on a wireless sensor network. Among other smartphone-based solutions: authors in \cite{li2020measuring} proposed to measure the plant growth characteristics in controlled environment agriculture; in \cite{golicz2020adapting}, authors analyzed the soil nutrients from 92 samples across Indonesia by using the smartphone as a portable reflectometer; in \cite{haro2019d}, authors presented an intuitive tool designed to help farmers to quickly understand their current and emergent drought and irrigation abstraction risks and support them during the decision-making. On the other hand, considering the recent advances in machine learning and smartphone hardware, the convolutional neural networks (CNN) have been integrated into smartphones to perform real-time classification of leaf diseases \cite{barman2020comparison, gulzar2020convolution, sanchez2019supervised, hinojosa2021fast}, which helps farmers to analyze the crop's health. Finally, the application of ICTs via the use of smartphones and its impact on the economy of the farmers has been previously studied in countries like China and India \cite{min2020does, parker2016enough}. However, few works report similar experiences from south American countries like Colombia.

In this paper, we present AgroTIC, a novel scientific-based mobile application that provides four specific solutions for the agricultural sector:
\begin{enumerate}
	\item To bring remote technical agricultural support required by farmers. It enables the possibility of diagnosis and plant disease identification in crops through image processing and machine learning techniques without further costs for farmers and technical advisors
	\item To acquire information related to crop, harvest, and production volumes, and to keep track of production management.
	\item To improve the food traceability from the crop to the market and provides a marketing platform to sell/buy agricultural products.
	\item To establish a communication channel between agronomists, merchants, and other farmers.
\end{enumerate}

In a general context, AgroTIC could provide economic development and reduction of costs to the farmer and create an open platform where farmers, merchants, and agronomists can coexist. Our proposed mobile App is a general solution and can be used by farmers in other Latin American countries similar to Colombia. This work significantly extends our previous conference paper \cite{conferenceAgroTIC}, which was published considerably before the AgroTIC project finished. Therefore, here we describe new software components we added to the App, \eg CNN for real-time classification of crop diseases and the merchant module, and present the final results we obtain with the farmers' community in Colombia.

This paper is organized as follows: in Section \ref{sec:related_works}, we reviewed related works focusing on improving agriculture processes using smartphones and other technology tools. In section \ref{sec:proposed_app}, we described our proposed mobile application for agriculture in detail. Specifically, we first defined the image processing and machine learning methods implemented in the AgroTIC app, and then we presented some technical details about the adopted software architecture. Finally, in Section \ref{sec:results}, we present and discuss some results obtained from using AgroTIC in the farming town of Simacota, located in the Santander Department, in Colombia.

\section{Related Works}
\label{sec:related_works}
The Food and Agriculture Organization of the United Nations (FAO) analysis concerning the future of global food and agriculture underscores the critical role of agriculture. The primary challenges to be addressed include sustainable improvement of agricultural productivity to meet growing demand, and the development of more efficient, inclusive, and resilient food systems. To confront these challenges, the FAO has committed to increasing agricultural productivity through innovation, incorporation, and use of ICT tools in traditional agriculture\cite{food, calicioglu2019future}. ICT has significant potential to enhance the efficiency, productivity, and sustainability of agriculture by facilitating information and knowledge sharing. Indeed, ICT can provide essential information on various crop variables, such as weather, water, soil, pests, and diseases, thus enabling farmers to make informed decisions in a timely manner\cite{ICT_2013}. Additionally, ICT can promote food traceability, accelerate operational speed, and level the playing field for small and large farmers. 

Numerous studies have reported on the advantages of ICT in agriculture. For instance, R. Smith and I. Meyer developed a policy for the role of ICT in agriculture in South Africa, considering several drivers of rural economic development\cite{ICT_SouthAfrica}. In India, A. G. Abishek, M. Bharathwaj, and L. Bhagyalakshmi employed web and mobile-based technologies for agricultural marketing activities\cite{ICT_India}. Similarly, Y. Zhang, L. Wang, and Y. Duan reviewed and analyzed agricultural information dissemination in China using ICTs\cite{ZHANG201617}. In Ghana, new ICT platforms for agricultural extension were developed and analyzed within their regional context\cite{MUNTHALI2018}.

\subsection{Internet and smartphone-based applications for Agriculture}

Leveraging Information and Communication Technology (ICT) and Internet of Things (IoT) tools can substantially enhance agricultural efficiency, productivity, and sustainability by offering real-time data, promoting knowledge exchange, and assisting in decision-making processes \cite{muangprathub2019iot, wishkerman2017application, chung2018smartphone, TRIPATHI2020183}. For example, Prasad et al. developed a mobile vision system using a smartphone to identify plant leaf diseases \cite{6953083}, Wu and Chang used GPS to detect pests and diseases in fields \cite{jagyasi2011event}, Sumriddetchkajorn used smartphone cameras to evaluate the level of rice leaves and recommend nitrogen application \cite{sumriddetchkajorn2013optics, sumriddetchkajorn2013mobile, intaravanne2012baikhao}, and Gomez et al. used smartphone cameras to obtain soil color information \cite{GOMEZROBLEDO2013200}. Other previous approaches include the use of GPS to propose a hybrid architecture for farmers that provides customized and reliable information in real-time \cite{jain2014hybrid} and the use of GPS, microphone, accelerometer, and gyroscope to remotely diagnose and treat plants using photos and videos \cite{saha2012development}. 

More recently, a smartphone and web application for optimal crop watering based on a wireless sensor network was suggested by the authors in \cite{muangprathub2019iot}. Other smartphone-based solutions include measuring plant growth characteristics in controlled environments \cite{li2020measuring}, using smartphones as portable reflectometers to analyze soil nutrients from 92 samples across Indonesia \cite{golicz2020adapting}, and providing an intuitive tool for farmers to quickly comprehend current and emerging drought and irrigation risks to aid decision-making \cite{haro2019d}. Additionally, recent advances in machine learning and smartphone technology have enabled the integration of convolutional neural networks (CNN) into smartphones for real-time classification of leaf diseases \cite{barman2020comparison, gulzar2020convolution}, assisting farmers in monitoring crop health. The impact of ICTs on farmers' economic well-being through smartphone use has been explored in countries like China and India \cite{min2020does, parker2016enough}. However, there is limited research on similar experiences in South American countries, such as Colombia.

\subsection{Internet of Things, Cloud Computing and Big Data to  Agriculture }

The Internet of Things (IoT) enables devices and sensors to send data through the internet in near-real time, resulting in growing interest and development in various fields such as industries, connected cars, and cities. In agriculture, this technology facilitates the development of sensor-based systems to monitor crop conditions, which can then be released to local farmers on small devices like smartphones \cite{IoT2018}. IoT integrates other ICT components such as cloud computing, big data, and data analytics. By 2024, it is projected that IoT device installations in the agricultural sector will increase from 40 million to 85 million, resulting in a significant increase in data generated from crops that must be stored and processed. Cloud computing becomes critical in this regard. However, as the data generated from IoT devices increases in speed, variability, and quantity, it can only be processed using big data platforms \cite{WOLFERT201769}. Several studies have developed IoT applications for field monitoring and automation, measuring parameters such as temperature, humidity, and soil moisture of crops, among others \cite{Mohanraj2016, PrashaunsaJ2018}.

\subsection{ICT in Colombian Agriculture}

Agriculture plays a crucial role in Colombia and other countries worldwide. Colombia, as reported by the Food and Agriculture Organization (FAO), is among the seven nations capable of satisfying the global demand for food due to its potential to expand crop production without resorting to forest clearing. Furthermore, Colombia is the third country worldwide with abundant water resources and climate diversity. However, despite these advantages, factors such as the increasing prevalence of diseases and pests, low productivity, limited technological innovation, and underutilization of information and communication technology (ICT) tools hinder the optimal use of available resources. To address these challenges, the Colombian government has launched several initiatives and public policies to promote technology adoption, increase farmers' productivity, and enhance their quality of life. One of these initiatives is the Strategic Plan of Science, Technology, and Innovation of the Agricultural Sector (PECTIA), developed in collaboration with the Ministry of Agriculture, Colciencias, Corpoica, and the Ministry of ICT became law in December 2017. PECTIA formulates objectives, strategies, and lines of action to improve the agricultural sector's productivity and competitiveness through ICT. In addition, the Ministry of Agriculture has developed platforms such as Agronet and Agronegocios, which offer technical assistance and information to beef cattle producers. Agronet also has an information network that enables producers to make informed decisions regarding the purchase and sale of agricultural inputs. On the other hand, most of the academic sector projects are supported by the Colombian government's research calls to promote solutions using ICT in the agricultural industry. Several universities in Colombia have also developed research projects focused on innovative solutions to the problems the agricultural sector faces. These projects seek to enhance productivity and competitiveness while improving the quality of life for farmers.

\section{Proposed Mobile Application for Agriculture}
\label{sec:proposed_app}

In general, the workflow of our proposed App is as follows: when farmers take a picture of their crop, the app performs image processing to extract features, such as vegetation indexes, and determine the health status of the plant. Furthermore, the algorithms behind the app detect and classify plant diseases using a convolutional neural network (CNN). Whenever the farmer requests a diagnosis by an agronomist, the AgroTIC platform sends the image processing and classification results to the agronomist to support the diagnosis. Then, the agronomist makes a decision and provides further technical assistance to the farmer via our app platform. 

\begin{figure}[t]
	\begin{center}
		\includegraphics[width=0.9\columnwidth]{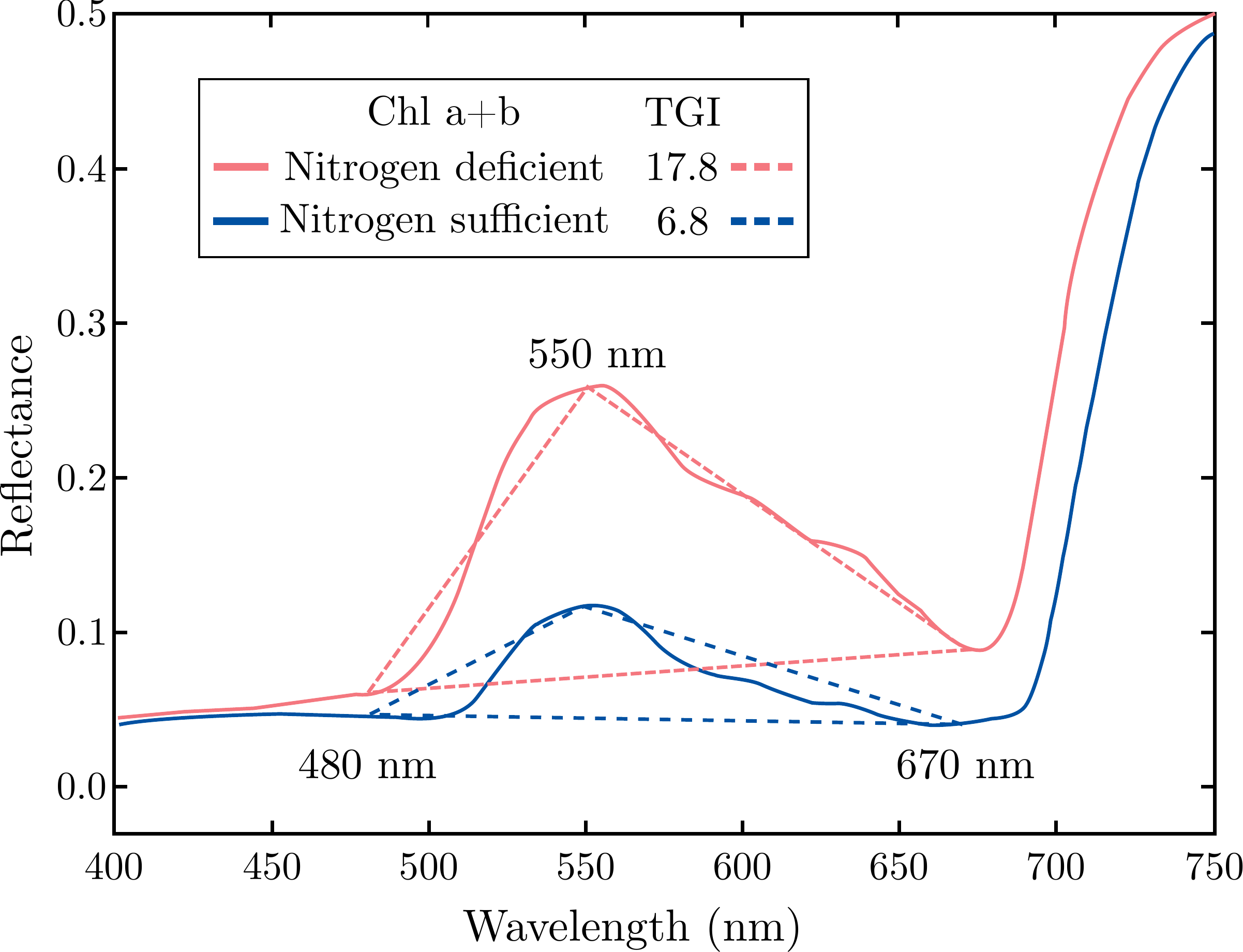}
		\caption{Examples of leaf spectral reflectance at chlorophyll a+b contents of 15 $\mu$g cm$^{-2}$ (nitrogen deficient) and 45 $\mu$g cm$^{-2}$ (nitrogen sufficient). The TGI is calculated from the area of a triangle defined by three points: $\rho_{480}$, $\rho_{550}$, and $\rho_{670}$, where $\rho_{\lambda}$ is the reflectance at wavelength $\lambda$.}
		\label{fig:TGI_example}
	\end{center}
\end{figure}
 
\subsection{Image processing and machine learning techniques}
\label{sec:VEI}

A vegetation index (VEI) is a transformation of spectral bands that allows obtaining a new image where the spectral properties of green plants are highlighted. VEIs typically support the diagnosis of the health of the plants in a crop \cite{bannari1995review}. The estimation of VEIs uses the principles of sunlight reflectance on the plants' leaves and are typically calculated from algebraic operations between different spectral bands, mainly for the spectral region of the visible (VIS of 400 to 680 nm) and the near-infrared region (NIR from 700 to 1000 nm) \cite{correa2017multiple}. Indeed, the total chlorophyll (Chl) content in leaves decreases in stress vegetation, changing the proportion of light-absorbing pigments and leading to a lower overall absorption due to lower Chl $a$ ad Chl $b$ concentrations at the leaf level. For instance, in Fig. \ref{fig:TGI_example}, Chl $a$ has a much higher absorption coefficient at 670 nm compared to 550 nm, so for a decrease in Chl content, the increase at 550 nm is larger (lower light absorption) than the increase at 670 nm.

Taking advantage of the RGB camera embedded in the smartphone, our AgroTIC App calculates two VEIs from the photos taken by the farmer. Formally, AgroTIC computes two vegetation indexes to assess the health of the plants defined as follows:\\

\mysection{Triangular Greenness Index (TGI) \cite{hunt2013visible}} TGI was developed to measure the chlorophyll content in the leaves. It is calculated by taking the area in the triangle that forms the spectral response at three wavelengths $\lambda$ of the VIS:
\begin{equation}
	TGI = -0.5\left[190(\rho_{670} - \rho_{550}) - 120(\rho_{670} - \rho_{480}) \right],
\end{equation}
where $\rho_{\lambda}$ is the spectral reflectance of the wavelengths $\lambda=670$nm, $550$nm, and $480$nm, respectively. Fig. 1 shows examples of TGI calculation to estimate if the leaf has a deficient or efficient nitrogen level. As shown, TGI is computed considering the area of a triangle; a low area indicates a sufficient nitrogen level (healthy leaf). It is calculated using the red, green, and blue bands of remote sensing imagery, which are sensitive to changes in chlorophyll content in vegetation. The TGI works by comparing the ratios of the green band to the red and blue bands, which helps to minimize the effects of soil and atmospheric variations. The TGI is widely used in agriculture, forestry, and environmental monitoring applications to detect vegetation changes over time and help guide management decisions. \\

\mysection{Green-Red Vegetation Index (GRVI) \cite{motohka2010applicability}} It is an indicator of phenological changes (plant life cycle) of the plants to allow detection in an early phase of leaf green-up. The variation among the relation of the green-red bands is the threshold used to diagnose Nitrogen problems in the plant. The GRVI index is defined as
\begin{equation}
	GRVI = \frac{\rho_{green}-\rho_{red}}{\rho_{green}+\rho_{red}},
\end{equation}
where $\rho_{green}$ and $\rho_{red}$ are the reflectance of green and red, respectively. Once the farmer takes a picture of a leaf of the plant, the RGB image is sent to AgroTIC's cloud computing platform, which estimates the VEIs. Then, the indices and the original image are sent to an agronomist, who will use them to support the diagnosis. Besides, the AgroTIC App also detects if the plant has a disease using a CNN and sends the results to the agronomist as additional information.

\begin{table}[t]
	\caption{Description of the convolutional neural network architecture implemented for the AgroTIC App.}
	\resizebox{\columnwidth}{!}{%
		\begin{tabular}{@{}cccccc@{}}
			\toprule
			Operation Layer & Number of Filters & Filter Size  & Stride      & Activation & Output Size               \\ \midrule
			2D Convolution  & 16                & $3 \times 3$ & $1\times 1$ & ReLU       & $224\times 224 \times 16$ \\
			2D Max Pooling  & 1                 & $2 \times 2$ & -           & -          & $112\times 112 \times 16$ \\
			2D Convolution  & 32                & $3 \times 3$ & $1\times 1$ & ReLU       & $112\times 112 \times 32$ \\
			2D Max Pooling  & 1                 & $2 \times 2$ & -           & -          & $56\times 56 \times 32$   \\
			2D Convolution  & 64                & $3 \times 3$ & $1\times 1$ & ReLU       & $56\times 56 \times 64$   \\
			2D Max Pooling  & 1                 & $2 \times 2$ & -           & -          & $28\times 28 \times 64$   \\
			Dropout         & -                 & -            & -           & -          & $28\times 28 \times 64$   \\
			Flatten         & -                 & -            & -           & -          & 50176                     \\
			Fully Connected & -                 & -            & -           & -          & 128                       \\
			Fully Connected & -                 & -            & -           & -          & 6                         \\ \bottomrule
	\end{tabular}}
	\label{tab:conv}
\end{table}

\begin{table}[t]
	\caption{Classification report for the testing dataset containing the five most common diseases found on citrus crops in Colombia.}
	\centering
	\resizebox{0.95\columnwidth}{!}{%
		\begin{tabular}{@{}cccccc@{}}
			\toprule
			& Alternaria & Acarus & Canker & Magnesium def. & Zinc def. \\ \midrule
			Precision & 0.94       & 0.97   & 0.84   & 0.98           & 0.92      \\
			Recall    & 0.94       & 0.97   & 0.91   & 0.95           & 0.88      \\
			F1-score  & 0.94       & 0.97   & 0.88   & 0.96           & 0.90      \\ \bottomrule
	\end{tabular}}
	\label{tab:report}
\end{table}

\begin{figure*}[t]
	\begin{center}
		\includegraphics[width=\linewidth]{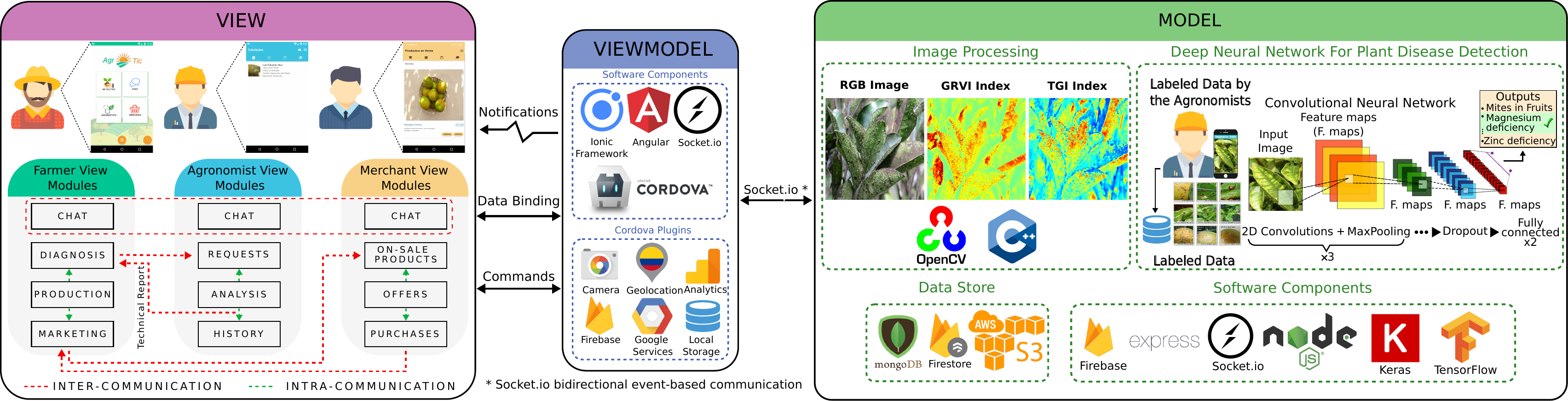}
		\caption{Model-View-ViewModel architectural pattern of the AgroTIC App.}
		\label{fig:MVVM_model}
	\end{center}
\end{figure*}

\begin{figure}[t]
	\begin{center}
		\includegraphics[width=0.95\columnwidth]{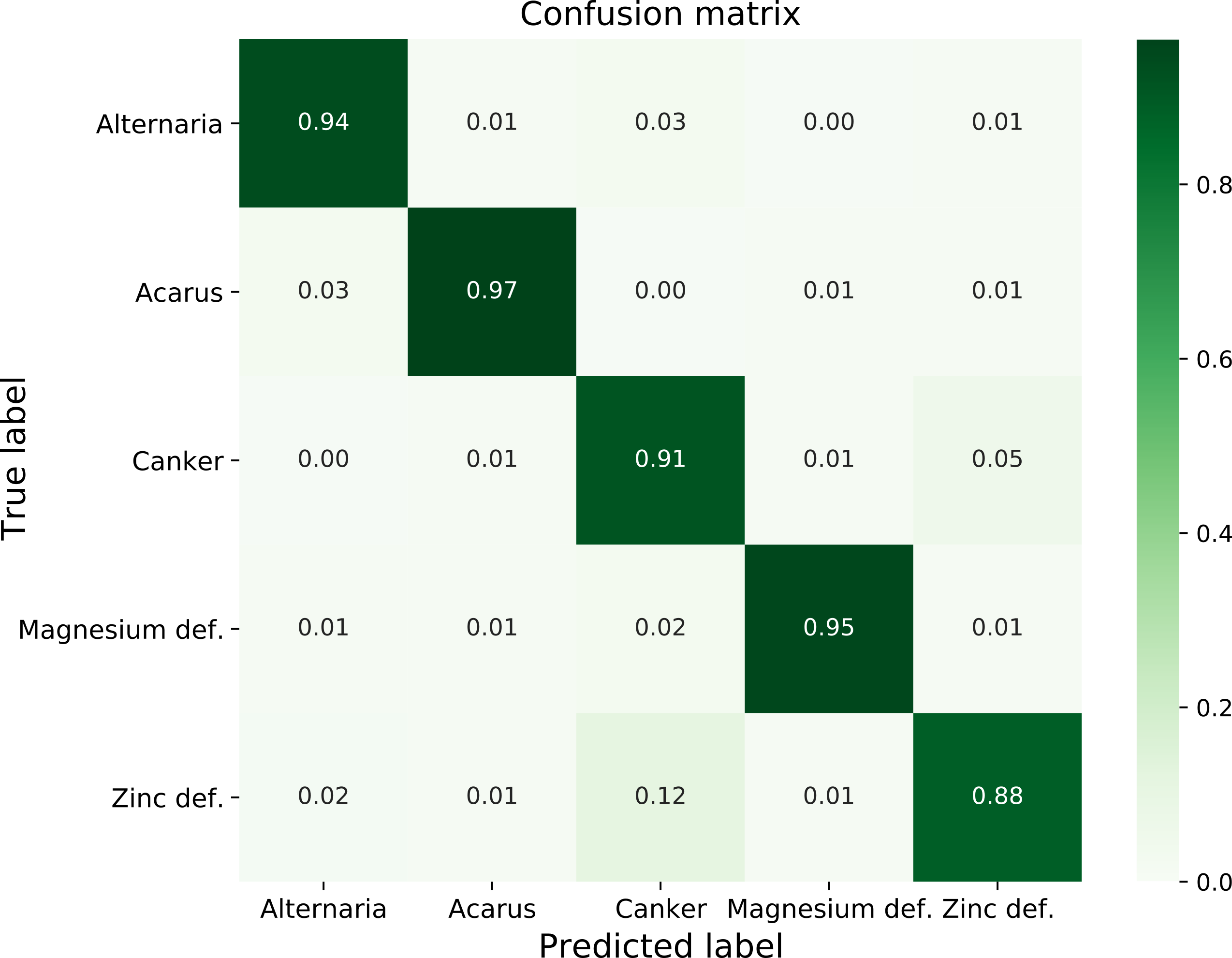}
		\caption{Classification results of the CNN implemented in AgroTIC App for citrus plant diseases classification. The overall accuracy of the model is $93\%$.}
		\label{fig:confusion_matrix}
	\end{center}
\end{figure}

\mysection{Convolutional Neural Network.} We trained a deep convolutional neural network (CNN) for citrus plant disease detection to further support agronomists in identifying unknown diseases in crop plants. We trained the first version of the model using 1250 images from the citrus plants, which have the five most common diseases found on citrus crops in Colombia: Alternaria, Acarus, Canker, Magnesium deficiency, and Zinc deficiency. In the set of images, we also include healthy photos to allow the model to detect them. We split the dataset into $80\%$ for training and $20\%$ for testing. Further, three data augmentation approaches were implemented to reduce overfitting: random horizontal flip, random rotation, and random zoom transformations \cite{shorten2019survey}. Table \ref{tab:conv} describes the proposed CNN implemented for AgroTIC, where the dropout layers also control overfitting. This model has $3,989,413$ trainable parameters, and we trained it for 200 epochs. Figure \ref{fig:confusion_matrix} and Table \ref{tab:report} show the performance of the trained model on the test set. The results show that the model can classify most diseases with more than $90\%$ of accuracy. The most challenging class to classify was Canker, with $88\%$ of accuracy. Analyzing the confusion matrix, we compute the overall accuracy of the model, obtaining $93\%$. It is worth mentioning that this CNN model will be continually trained using the technical responses and newly labeled data provided by agronomists within AgroTIC App. In this way, AgroTIC can be seen as a collaborative platform where agronomists help to improve the CNN disease detection model.


\subsection{Software Architecture}
\label{sec:proposed_mobileApp}

The vegetation indexes calculation and the CNN described in the previous Section are supported in a technological infrastructure based on a cloud computing platform responsible for managing requests from farmer users to the ICT platform. Specifically, AgroTIC is a hybrid mobile app developed using Ionic Framework \cite{IonicFrameworkwithAngularforHybridAppDevelopment}, an open-source UI toolkit for building performant, high-quality mobile apps using web technologies (HTML, CSS, and JavaScript). AgroTIC App follows the Model-View-ViewModel (MVVM) architectural pattern \cite{anderson2012model}, which is the standard architecture used by Ionic-based mobile applications. Using the MVVM pattern, the UI of the application and the underlying presentation and business logic is separated into three classes as shown in Fig. \ref{fig:MVVM_model}: \textbf{the View}, which encapsulates the UI and UI-logic; \textbf{the View Model}, which encapsulates presentation logic and state; and \textbf{the Model}, which encapsulates the application's business logic and data.

\subsection{The View}
AgroTIC comprises three view designs depending on the user role: farmer, agronomist, and merchant (see Fig. \ref{fig:MVVM_model}). Each view design has four modules allowing users to perform their role-defined activities and interact with others.

\mysection{The Farmer view} contains the modules: Chat, Diagnosis,  Production, and Marketing. The Chat module allows farmers to establish a communication channel with other farmers and agronomists. With this communication network, farmers can share experiences and consult other farmers about concerns or common crop problems. The Diagnosis module allows farmers to take pictures of leaves from a plant with an unknown disease. Then, it runs the vegetation index algorithm and the CNN on the acquired images to estimate the plant's overall health and the disease or condition, respectively. Finally, the Diagnosis module sent the results to an agronomist to further analysis. The Production module allows farmers to keep track of their lands and crops, and this information is shared among the Diagnosis and Marketing Modules. Finally, the Marketing Module will enable farmers to offer their products, start and keep track of their auction, and establish direct contact with the end buyers. 

\mysection{The Agronomist view} contains the modules: Chat, Requests, Analysis, and History. The Request module shows a list of analysis requests of diseased plants made by farmers. For each request, the agronomist can view the vegetation indices of the diseased plant, which were obtained in the cloud computing platform using image processing. Once selecting an analysis request, the agronomists use the Analysis module to prepare a diagnostic report sent back to the farmer, and a log of the process is saved in the History. 

\mysection{The Merchant view} contains the modules: Chat, On-sale Products, Offers, and Purchases. After a farmer publishes the products using the Marketing module, such products are listed in the On-Sale Products module of the Merchant, allowing them to make an initial offer and see the detailed description of the product that contains photos, product quantity, and production information. The Offers module allows the merchant to keep track of a particular product, notifies if their current offer has been overcome, and enables them to make a new offer. Finally, when the auction time ends, if the merchant wins, the contact information is sent back to the farmer, and the purchase log is saved in the Purchases module. See red dashed lines in Fig. \ref{fig:MVVM_model}.

\subsection{The View Model}

The AgroTIC View model provides methods, commands, and other functions to update the View and manipulate the Model. The Ionic Framework handles all the user experience of the App, such as controls, interactions, gestures, and animations. The notifications sending and event handling within the View Model, as well as the mobile-optimized web technology-based components, are provided by the Ionic Framework as well. Besides, as an Ionic App, most of the methods for updating the View, such as the data binding functionality, are provided by the Angular Framework. On the other hand, the queries, validations, and update requests to the Model are implemented using Socket.io \cite{rai2013socket} (client-side). Indeed, Socket.io enables real-time, bi-directional communication between the View Model, which uses the client-side library, and the Model, which uses the server-side library for Node JS, see next subsection. Finally, to manipulate the smartphone hardware, such as sensors, cameras, and GPS, and to store local data, AgroTIC relies on the hybrid mobile application development framework Apache Cordova \cite{fermoso2009phonegap}. Apache Cordova is a hybrid mobile application development framework that enables software programmers to build applications for mobile devices using CSS3, HTML5, and JavaScript instead of relying on platform-specific APIs like those in Android and iOS.


\subsection{The Model}
\label{subsec:themodel}

The AgroTIC Model handles all the application's business logic and data by a cloud computing platform set up with a web service using NodeJS, Express, and Socket.io (server-side) for bi-directional communication, and data storage services, including MongoDB, Firestore, and Amazon Simple Storage Service (S3). The cloud computing platform incorporates an image analysis and processing system that implements state-of-the-art algorithms developed in C++ from the open-source computer vision library OpenCV. Such algorithms are encapsulated in a computing service application, whose inputs and outputs are handled in the NodeJS server through WebSockets and HTTP requests. Here we encapsulate the vegetation indexes calculation and the trained CNN for plant disease detection.

\begin{figure*}[t]
	\begin{center}
		\includegraphics[width=\linewidth]{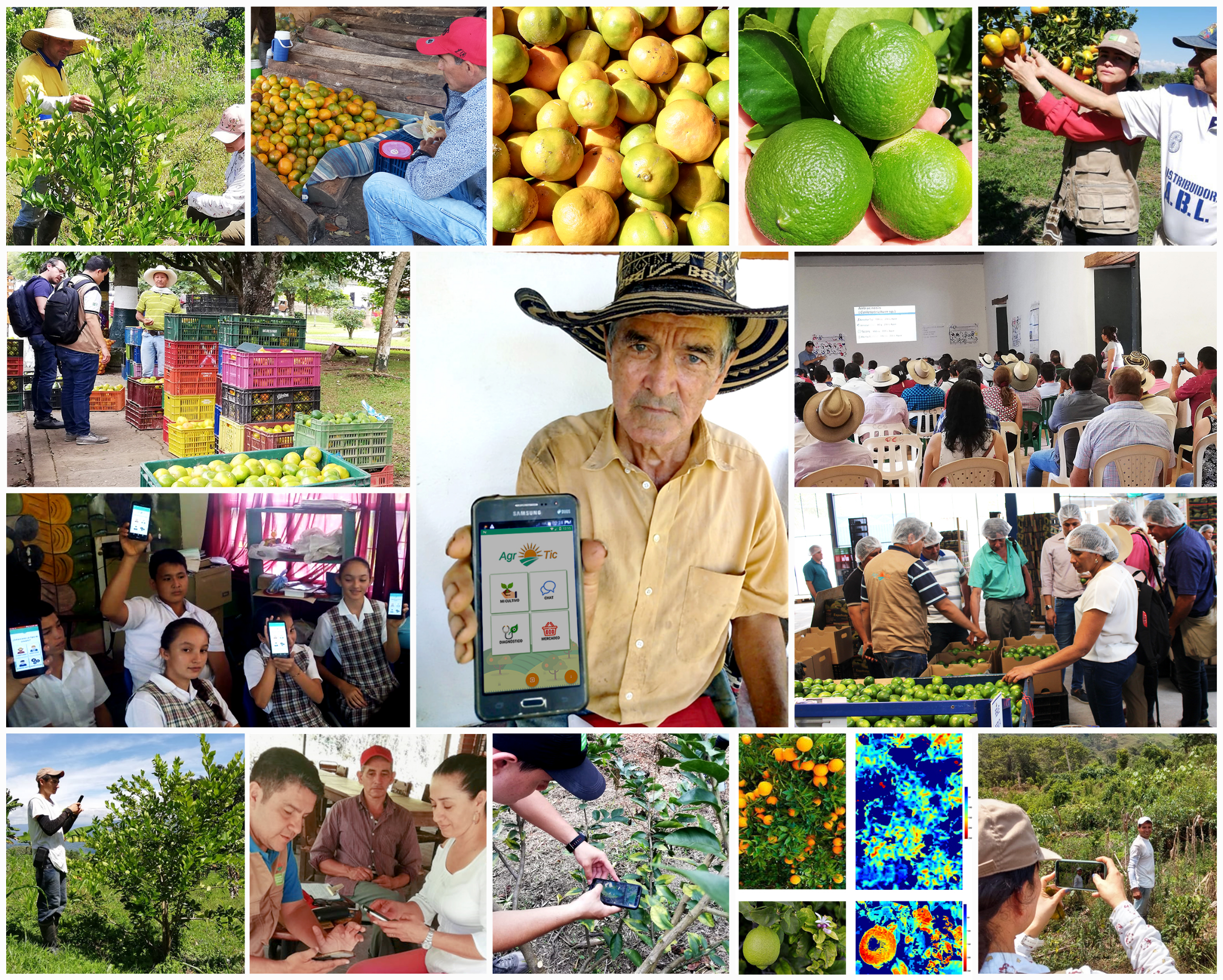}
		\caption{Activities aimed to spread AgroTIC App usage among farmers and locals. These activities included workshops in the Simacota downtown, farms, and high schools.}
		\label{fig:AgroTICActivities}
	\end{center}
\end{figure*}

\section{Results: AgroTIC in Colombian farming Communities}
\label{sec:results}


\mysection{The project.} In this section, we report some results obtained from AgroTIC in the farming town of Simacota, located in the Santander department in Colombia. Here we report the results obtained during the main project length (eleven months). In general, this project was developed (excluding the farmers' representatives) by sixteen people, including Ph.D. project director (1), agronomists (2), software developers (4), ICT experts (1), Colombian government staff (4), logistic staff (1), marketing staff (1), and social media staff (2). AgroTIC impacted a community of 200 farmers who grow citrus fruit (orange, tangerine, and Tahiti lime).

\begin{figure}[h!]
	\begin{center}
		\includegraphics[width=0.9\columnwidth]{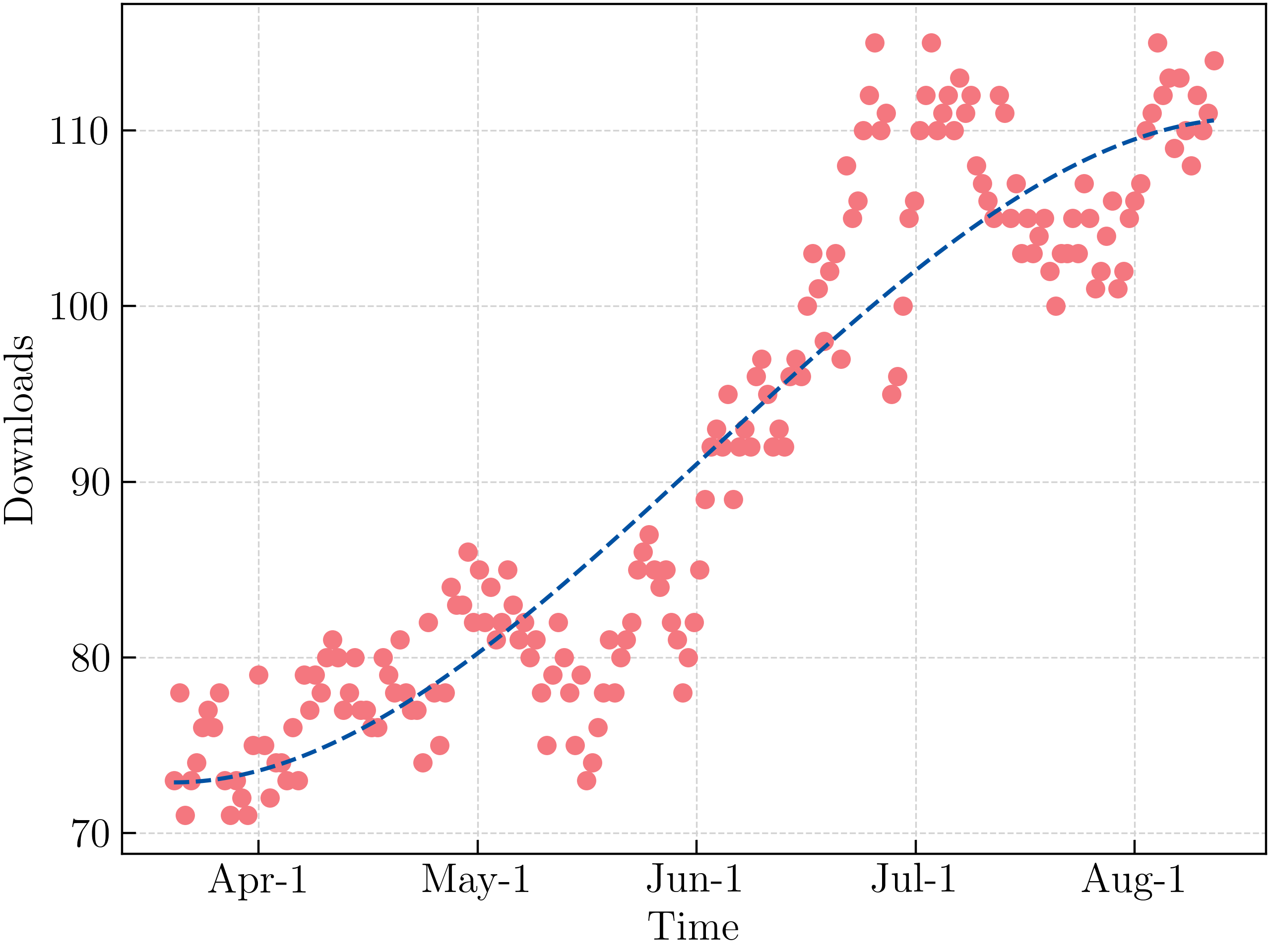}\vspace{-0.1in}
		\caption{Number of downloads per day in a three-month time window. The trend line is estimated using local regression (loss). The shaded area corresponds to the loss 95\% confidence interval estimate.}\vspace{-0.2in}
		\label{fig:AgroTICStats}
	\end{center}
\end{figure}

\mysection{Activities with the community.} Once the App was ready for download, we designed and implemented several strategies to spread its use among locals. First, we organized three main workshops in Simacota town (Santander, Colombia). One of these workshops showed the locals how ICTs could improve their farm's productivity. About 100 people attended the event, between farmers and locals. The second workshop intended to expose the locals to the new App AgroTIC, showing its main features and critical use cases that local farmers could exploit to improve farming activities and profit. In the last workshop, we showed farmers the Best Practice Guidelines (BPG) for sustainable agriculture and value chains. In this workshop, we asked locals what are their common agricultural practices and then compared them with BPGs. With the agronomists, we visited more than 80 farms from 6 villages in Simacota's rural area. In those visits, we recommended the implementation of BPGs related to plant disease control, soil composition monitoring, and efficient management of waste derived from soil enrichment practices. After the initial workshops, we visited the downtown, schools, and farms close to Simacota to keep spreading the App usage. During those visits, we answered questions about the App's main features, showed locals tutorials on App usage, and spread BPGs among locals and individual farms. Figure \ref{fig:AgroTICActivities} shows some developed activities. Figure \ref{fig:AgroTICStats} shows the number of downloads in a three months time window. By the time this paper was written, downloads had increased about 10 percent compared to the first day the application was available for download.

\mysection{Main results within the community.} At the end of the project, 74 farmers affirmed a reduction in crop losses thanks to the support they obtained from agronomists through AgroTIC, hence increasing production. Also, during this study, 102 farmers connected with 12 merchants and offered 65 products using the AgroTIC platform. Table \ref{tab:usage} reports the AgroTIC usage frequency since the application was launched. As shown, a total of 171 users, among farmers (146), Agronomists (9), and merchants (12), registered in the App; they started 171 chats and exchanged 1350 messages, taking 38 pictures (samples) from 275 registered crops to being processed and send to agronomists, and offered 65 products using the AgroTIC market platform.

\begin{table}[t]
	\setlength{\tabcolsep}{4pt}
	\resizebox{\columnwidth}{!}{
		\begin{tabular}{@{}ccccccccc@{}}
			\cmidrule(r){1-3}
			\multicolumn{3}{c}{Total Users: 171} &       &         &          &          &       &       \\ \midrule
			Farmers  & Agronomists  & Merchants  & Chats & Samples & Products & Messages & Farms & Crops \\
			146      & 9            & 12         & 171   & 38      & 65       & 1350     & 80   & 275   \\ \bottomrule
	\end{tabular}}
	\caption{Usage frequency of the different App features. }
	\label{tab:usage}
\end{table}

\begin{figure}[h!]
	\begin{center}
		\includegraphics[width=\columnwidth]{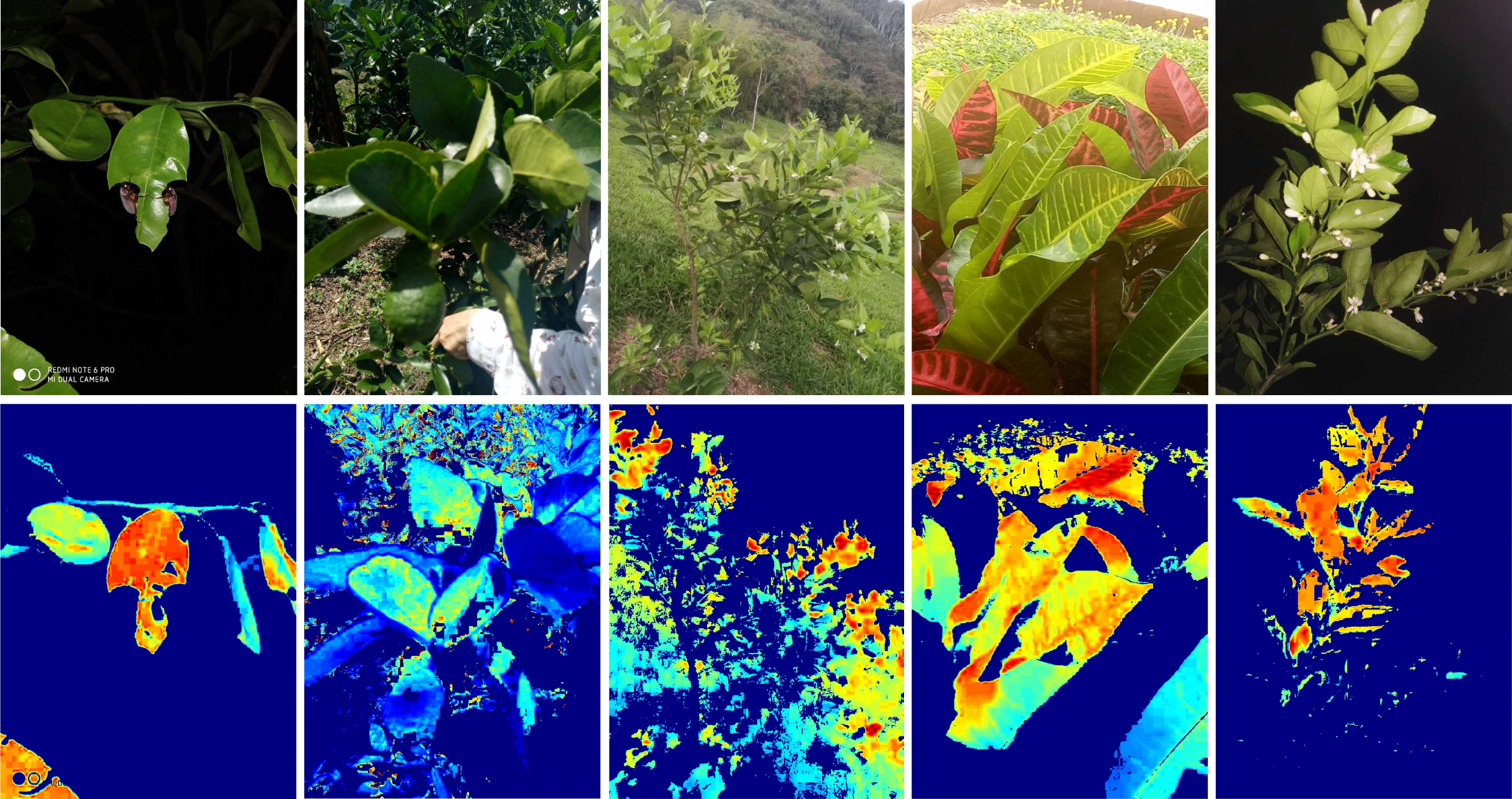}\vspace{-0.1in}
		\caption{Photos of crops acquired by smartphone cameras from farmers which use our AgroTIC (top) App and Triangular Greenness Index (bottom).}\vspace{-0.2in}
		\label{fig:TGI}
	\end{center}
\end{figure}

\vspace{0.2cm}
\mysection{Qualitative results}. Here we show some qualitative results of the vegetation indices calculated by the AgroTIC platform. These images were acquired by the farmer during the study time of this project and sent to our platform. Then, using our classification network and the feedback given by the agronomist, the farmers successfully improved their crops by following the guidelines provided in the AgroTIC app. Specifically, fig. \ref{fig:TGI} shows some original photos of crops acquired with smartphone cameras by farmers (above) and their respective calculations of the Triangular Greenness Index (TGI) (bottom). In the images in the bottom row, a high TGI value, represented in blue, indicates healthy and productive vegetation with high chlorophyll content, while a low TGI value, in red, indicates stressed vegetation. As we mentioned in section \ref{sec:VEI}, the TGI is widely used in environmental, agricultural, and forestry monitoring applications to detect changes in vegetation over time and help guide management decisions. Therefore, the image resulting from this step supports the diagnosis made by the agronomist.

\section{Conclusions and Future Plans}
\label{sec:conclusion}

This paper presented one of the first ICT tools developed for the Colombian agricultural sector, supporting farmers in their agricultural practices. With the AgroTIC project, we reduce the digital gap among the small farming community in Simacota, Santander, using smartphones. AgroTIC's impact is evidenced by the increasing number of downloads and App usage among farmers in traditional agriculture activities. Use cases of the App range from crop disease diagnosis through image processing and vegetation index analysis to marketing support of citrus products. In the near future, we expect to impact more Colombian farming communities that grow different crops and include more functionalities to AgroTIC App to improve farming productivity. We further expect that the AgroTIC app can be used in other countries where there still exists a digital gap in their farming communities.



\section*{Acknowledgments}
\label{sec:acknowledgments}

The AgroTIC project was supported by the Colombian Ministry of Science, Technology and Innovation (MinCiencias) and the Ministry of Information Technologies and Communications (MINTIC).

{\small
\bibliographystyle{ieee_fullname}
\bibliography{egbib}
}

\end{document}